# On Transition Metal Catalyzed Reduction of N-nitrosodimethlamine

Jun Zhou[1*], Min Wang[1], Junhua Tian[1], Zhun Zhao[1]
[1]Chinese-American Chemical Society, Houston, TX 77054
*Corresponding author, zhoujun06342@163.com

**Abstract:**

This report provides a critical review on "Metal-Catalyzed Reduction of N-Nitrosodimethylamine with Hydrogen in Water", by Davie et al.[1] N-nitrosodimethlamine (NDMA) is a contaminant in drinking and ground water which is difficult to remove by conventional physical methods, such as air stripping. Based on the reported robust capability of metal based powder shaped catalysts in hydrogen reduction, several monometallic and bimetallic catalyst are studied in this paper on the reduction of NDMA with hydrogen. Two kinds of kinetics, metal weight normalized and surface area normalized, are compared between each catalyst in terms of pseudo-first order reaction rate. Palladium, copper enhanced palladium and nickel are found to be very efficient in NDMA reduction, with half-lives on the order of hours per 10 mg/l catalyst metal. Preliminary LC-MS data and carbon balance showed no intermediates. Finally, a simple hydrogen and NMDA surface activated reaction mechanism is proposed by the author for palladium and nickel.

**Background:**

NDMA (Figure 1), also known as dimethylnitrosamine (DMN), is a semi-volatile organic chemical that is highly toxic and is a suspected human carcinogen. The US Environmental Protection Agency has determined that the maximum admissible concentration of NDMA in drinking water is 7 ng $L^{-1}$.

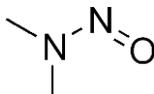

Figure 1. Chemical structure of NDMA [2]

The pollution source and hazards of NDMA, are listed as below:

1. Pollution sources

    i) Dietary sources: in various meat and cured meat products (600 to 1,000 ng/kg in fried pork bacon), fish and fish products, beer (50 to 5,900 ng/kg), milk (90 to 100 ng/L), cheese, soybean oil, canned fruit, and apple brandy. In the body, NDMA is formed when acidic conditions in the stomach catalyze the reaction between nitrite and dimethylamine (DMA).[3]

    ii) Non-dietary sources: unsymmetrical dimethylhydrazine (UDMH, a rocket fuel component which used NDMA during synthesis), cutting oils, tobacco smoke, herbicides, pesticides, rubber products, and drugs formulated with aminopyrine. Recently discovered sources of NDMA precursors include carpet dyes, dithiocarbamates and methyl-dithiocarbamates from circuit board manufacturers.

2. Hazard
   Carcinogen (EPA ranked); chronic health effects including liver cancer, lung cancer, non-cancerous liver damage, internal bleeding, and neurological damage.

Elevated levels at many wells throughout California and Canada have alarmed the potential threat that NDMA might have on human beings and ecosystem. Thus, a cost-efficient way of treating NDMA polluted water is urgent. In this part, several traditional and current methods of disposal are compared with respect to their efficiency, by-product, cost, and feasibility.

Table 1. Summary of methods to treat N-nitrosodimethlamine [4]

| **Methods** | | efficiency | By-product | cost | feasibility |
|---|---|---|---|---|---|
| Adsorption | Volatilization, air stripping, granular activated carbon adsorption | Low (small Henry constant and octanol-water partition coefficient) | no | low | low |
| Advanced oxidation | Hydroxyl radicals from hydrogen peroxide or ozone using UV radiation | Low (limited by radical scavengers) | Slight | Low | low |
| Biological degradation | Aerobic bacteria | mixed (long half-lives: 12-55 days) | Slight | - | medium |
| Reverse osmosis | Reverse osmosis membrane | Medium (50% removal) | No | - | medium |
| Direct UV photolysis | UV radiation | Medium (50% removal with 1 day residential time) | - | high | Currently used for small scale |
| Reductive catalysis | Pd, Cu-Pd, Fe, Ni, Fe-Ni, Mn [1, 5] | Relatively high (half-lives on the order of hours) | DMA and ammonium | - | promising |

For each different method, the major drawbacks are:

1. Adsorption: When saturated in water, the partial pressure of NDMA is calculated as: $2.63 \times 10^{-4}$ atm $M^{-1} \times 3.914$ M=$1 \times 10^{-3}$ atm, which is so low that volatilization or air stripping will not significantly affect. Also, due to a very low partition coefficient in other medium, such as soil and granular activated carbon (GAC), it is also inefficient to be absorbed by GAC or soil.
2. Advanced oxidation: Radical scavengers in polluted water can limit the concentration of radicals required by practical operation.
3. Biological degradation: No significant loss of NDMA is observed during field test at the Rocky Mountain Arsenal. Usually it takes days to months to approach 50% removal. Besides, the complicated interactions during bio-degrading have not been clarified yet.

4. Reverse osmosis: Only 50% removal is achieved, and post-reverse osmosis treatment is necessitated.
  5. Direct UV photolysis: Cost-effective in small scale treatment of NDMA, and is currently used method. However, for large scales, it is no longer cost-effective.
  6. Reductive catalysis: have half-lives on the order of hours, and are considered as very promising alternatives for Direct UV photolysis. However, the poisoning and regeneration issues are important, and the reaction pathway needs to be clarified.

The objectives of this paper are stated clearly in the introduction part, which are:
  1. Determine the activity of various metal catalysts for NDMA reduction.
  2. Explore the impact of copper enhancement of supported Pd catalyst.
  3. Quantify reaction intermediates and catalysts.

**Materials and Methods:**

Table 2. The metal catalysts and their properties

|  | **Pd** | **Pd-Cu** | **Ni** | **Fe** | **Fe-Ni** | **Cu** | **Mn** |
|---|---|---|---|---|---|---|---|
| **Composition** | 1wt% Pd/ϒ-Al$_2$O$_3$, Sigma Aldrich | 1wt% Pd-0.3wt% Cu / ϒ-Al$_2$O$_3$, Alfa Aesar (Ward Hill, MA) | 99.99%, Sigma Aldrich | 99.99% zerovalent Fe, Alfa Aesar (Ward Hill, MA) | 42wt% Ni/Fe, Alfa Aesar (Ward Hill, MA) | 99+%, Sigma Aldrich | 99+%, Sigma Aldrich |
| **BET surface area (m$^2$ g$^{-1}$)** | 153.3 | 153.3 | 1.5 | 0.5 | 0.3 | 0.5 | ND$_b$ |
| **pore volume (cm$^3$ g$^{-1}$)** | 0.77 | 0.45 | 0.01 | 0.001 | 0.003 | 0.002 | 0.001 |
| **pore size distribution (nm)** | 39±9 | 18±6 | 22±8 | 22±8 | 22±8 | 33±13 | 22±10 |
| **metal content (%ww$^{-1}$)** | 1 | 1, 0.3 | 99.99 | 99 | 58, 42 | 99 | 99 |
| **metal dispersion (%)** | 45 | 45 | - | - | - | - | - |

Table 3. Chemicals in use in the experiment

| 18.2 MΩ cm$^{-1}$ Milli-Q water | Synergy 185 Millipore with Simpak2 purifying system (Millipore; Billerica, MA). |
|---|---|
| NDMA/ methanol | Supelco; Bellefonte, PA |
| deuterated NDMA (NDMA-d$_6$)/ methanol | Supelco; Bellefonte, PA |
| H$_2$, 99.995% | Praxair (Danbury, Connecticut) |
| DMA | Sigma-Aldrich; St. Louis, MO |

Table 4. Reaction conditions

| | |
|---|---|
| **temperature (K)** | 294±1 |
| **headspace hydrogen pressure (psig)** | 8 |
| **[H2]$_{(aq)}$ ($\mu$M)** | 800 |
| **volume (L)** | 2.2 |
| **aqueous/vapor (L/L)** | 2.0/0.2 |
| **pH** | 5.6 |
| **mixing speed (rpm)** | 600 |
| **C$_{me}$ (mg L$^{-1}$)** | 1-20 |
| **[NDMA]$_0$ ($\mu$g L$^{-1}$)** | 100 |

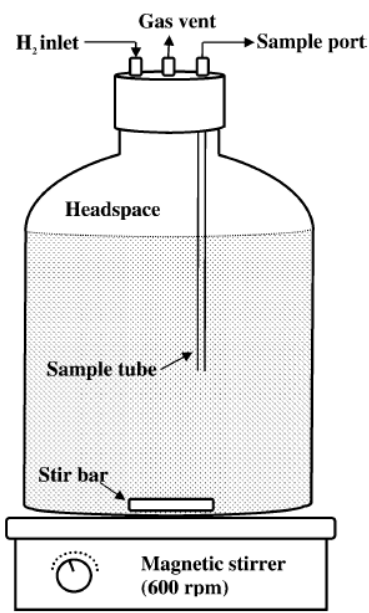

Figure 2. Reactor set-up (adapted from [1])

*Reaction procedure:*
Reaction procedure in this paper is described as in the following scheme.

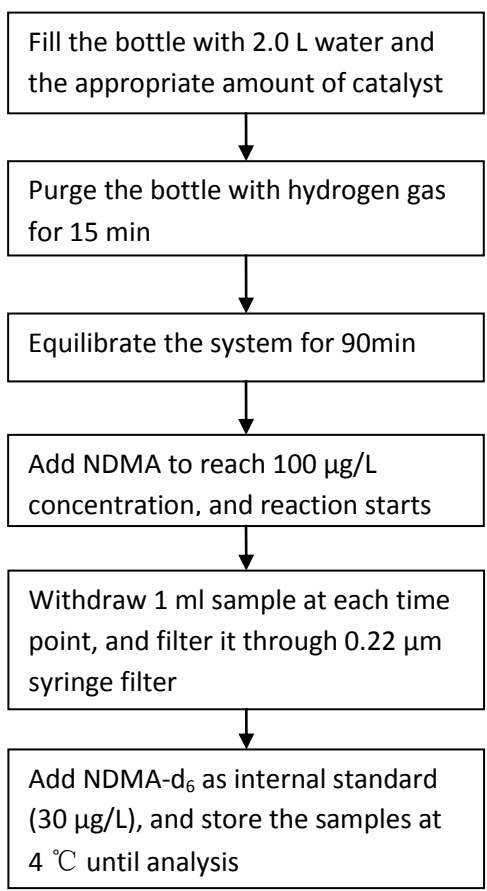

Scheme 1. Reaction procedure

*Analytical methods:*

Table 5. Detection of NDMA and DMA using LC-MS

| LC/MS | API 3000 high-pressure LC-MS/MS, Applied Biosystems; Foster City, CA |
|---|---|
| **Column** | A C-18 column (5 ṁ particles, 50×2.1 mm, Higgins Analytical; Mountain View, CA) |
| **Injection amount** | 50 μL |
| **Mobile phase** | Mixture of 2 mM ammonium acetate and methanol<br>1. 10% methanol<br>2. Increase linearly to 40% methanol over 6 min<br>3. Increase linearly to 100% methanol over 6 min<br>4. Remain at 100% methanol for 4 min |
| **Flow rate** | 150 $\mu$L min$^{-1}$ |
| **NDMA MS peak** | 75/43 transition |
| **NDMA-d$_6$ MS peak** | 81/46 transition |
| **DMA MS peak** | 46 amu peak |

Detection of ammonium

1. Mix the sample with ammonium cyanurate to buffer the solution to pH>10.
2. Prepare the standard curve by detecting the known concentration ammonium solutions.
3. Detect the concentration of ammonium using photometric probe.

**Reaction kinetics:**

$$-\frac{1}{C_{me}}\frac{dC_{NDMA}}{dt} = k'_{obs}C_{NDMA} \qquad (1)$$

$$-\frac{1}{SA_{me}}\frac{dC_{NDMA}}{dt} = k'_{obs}C_{NDMA} \qquad (2)$$

The author proposed a pseudo-first-order reaction rate model expressed above, where $k'_{obs}$ is the observed reaction rate constant normalized by active metal concentration or metal surface area per volume for weight and surface area normalization, respectively.

To consolidate the assumption that the weight of catalysts have little or no effect on the $k'_{obs}$, the author conducted a series of reactions on their relationship.

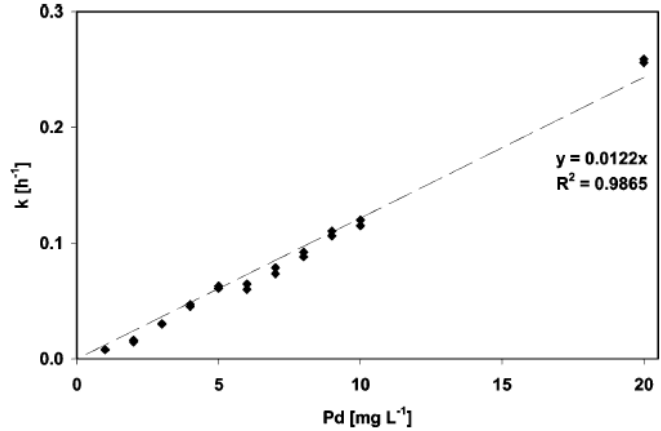

Figure 3. Reaction rate as a function of Pd loading in the reactor (adapted from [1])

The figure shows good linearity of observed pseudo-first-order rate constant and catalyst weight for all the cases. Thus, $k'_{obs}$ is independent of metal concentration or metal surface area per volume.

**Results and discussion:**

Reaction kinetics

Table 6. Pseudo-first-order reaction kinetics (adapted from [1])

| catalyst | catalyst composition | metal weight loading normalized pseudo-first-order rate[a] [L $g_{me}^{-1}$ $h^{-1}$] | metal surface area normalized pseudo-first-order rate[b] [L $m_{me}^{-2}$ $h^{-1}$] | half-life[c] [h] |
|---|---|---|---|---|
| Pd | 1% Pd | 11.5 ± 0.9 | 0.17 ± 0.01 | 6.0 ± 0.4 |
| Pd–Cu | 1% Pd–0.3% Cu | 66.5 ± 7.4 | 1.05 ± 0.12 | 1.0 ± 0.1 |
| Ni | 99% Ni | 8.3 ± 2.9 | 5.53 ± 1.93 | 8.4 ± 2.2 |
| Fe | 99% Fe | 0.13 ± 0.09 | 0.26 ± 0.18 | 533 ± 218 |
| Fe–Ni | 42% Ni–58% Fe | 0.65 ± 0.01 | 2.17 ± 0.03 | 107 ± 2 |
| Mn | 99% Mn | 0.07 ± 0.02 | | 990 ± 220 |
| Cu | 99.99% Cu | 0 | | |
| Al$_2$O$_3$ | 99.99% Al$_2$O$_3$ | 0 | | |

[a] Pseudo-first-order rates are normalized by active metal weight (n > 8). [b] Pseudo-first-order rates are normalized by metal surface area (n > 8). [c] Calculated half-lives for 10 mg $L^{-1}$ active metal.

Two kinds of kinetics are investigated by the author, metal weight normalized kinetics and surface area normalized kinetics. The data is summarized in the above table, and conclusion on the author's analysis can be drawn as:

1. control group: ϒ-Al$_2$O$_3$, Cu
No reaction is observed for ϒ-Al$_2$O$_3$, indicating that it is inert during hydrogenation.

2. Fe, Fe-Ni, Mn group:
As indicated by the author, metals in this group generate atomic hydrogen from water, which is then harnessed in the hydrogenation. This additional step slows down the reaction rate, compared with the other group harnessing dissolved hydrogen.
Ni enhances Fe's reaction activity by 5-fold in the study. The mechanism is not specified.

3. Pd, Pd-Cu group:
Pd based metals are shown to be very active in the hydrodechlorination reactions.[6-22] Herein Pd based metals has a relatively greater reaction activity in metal weight normalized kinetics than Fe based metals. Since Cu can enhance the reaction activity of Pd by 6-fold in the study and Cu shows no activity, the author proposed the mechanism that NDMA can be rapidly activated on the Cu surface, and the activated NDMA is then reduced by the atomic hydrogen activated by Pd.

4. Ni
Ni was tested with dihydrogen as the electron donor, as indicated by the author. And it exhibits a rather high metal weight normalized reaction rate (8.3±2.9 L/$g_{me}$/h) and the highest surface normalized reaction rate (5.53±1.93 L/$m_{me}^2$/h).

The surface area normalized reaction rates are higher for Fe, Fe-Ni, Ni group than Pd, Pd-Cu group, indicating that Fe, Fe-Ni and Ni take more advantage of their surfaces, making them promising for application with respect to high surface efficacy.

**Reduction products:**

The reduction products of NDMA by Pd, Pd-Cu, and Ni are DMA and ammonium (proposed). No peaks other than NDMA and DMA observed in LC-MS indicate that there are no intermediates during the reaction. Although ammonium has not been detected in the study, it is thought to be the expected products, and the related study is still in investigation. For Fe and Fe-Ni metals, although no experiment has been conducted, according to previous study, their desired reduction products should also be DMA and ammonium.

The concentration profiles of NDMA and DMA are drawn in the figure below. It shows good fit of the pseudo first-order model, and the C-balance profile indicates that no intermediates during reaction.

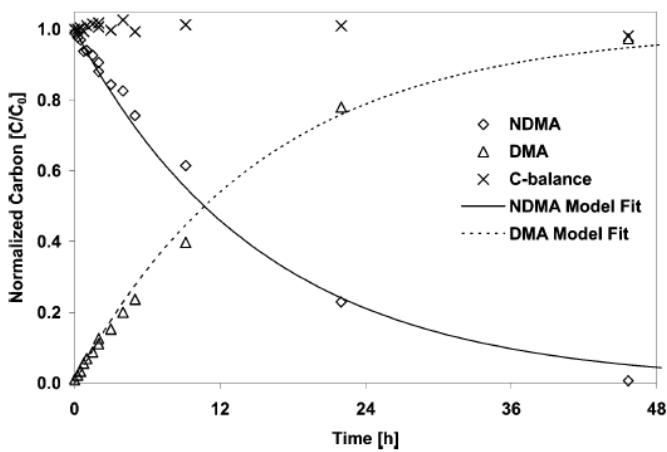

Figure 4. Reaction profiles using 1% Pd on $Al_2O_3$ (adapted from [1])

**Hypothesized mechanism**

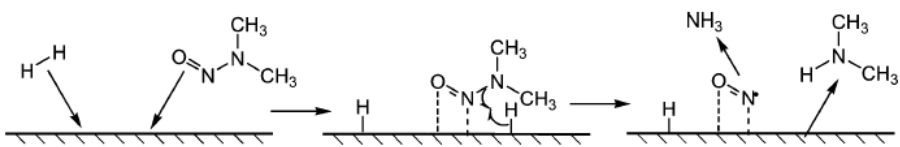

Scheme 2. Proposed reaction mechanism (adapted from [1])

For Pd and Ni, as proposed by the author, the reaction undergoes the following steps:

1. NMDA and dihydrogen diffuse to active metal sites. And diffusion is fast enough to be negligible.
2. NDMA and $H_2$ are adsorbed and activated on the surface. Atomic H is very active and mobile.
3. N-N bond is cleaved, the amine group is reduced to DMA.
4. Nitrosyl radical is reduced to ammonium.
5. Ammonium and DMA detach from the surface and diffuse away.

For Cu enhanced Pd, the proposed mechanism is quite the same, but Cu is hypothesized to provide rapid activation of NDMA.

**Critical review:**

The introduction part of this paper is rather comprehensive. The sources and hazards of NDMA have been pointed out, and all kinds of methods of treating NDMA, both currently using and promising, are introduced. However, it is better for the author to summarize those methods in a more uniform manner with respect to advantages and disadvantages as the table above. In this way, a much clearer landscape of this field will be presented, and the potentials of the metal catalysts in the study will be clarified through comparison.

In the evaluation of reaction kinetics part, the author pointed out that "hydrogen was not included in the rate expression because the activation energy for dihydrogen chemisorption onto noble metals is close to zero". This is not the whole reason, because activation is only part of the process that hydrogen involves. The reduction of NDMA also needs hydrogen where the concentration of hydrogen takes into play. The major point is that the saturated concentration for $H_2$ (800 μM) is much higher than NDMA (1.35 μM). Thus, the original 2$^{nd}$ order reaction for $H_2$ and NDMA reduces to a 1$^{st}$ order one because [$H_2$] is a constant, which merges into the $k_{obs}$. However, this introduces another problem, which is: for those not using dissolved hydrogen, Fe, Fe-Ni, Mn, the concentration of hydrogen generated by metal itself is not specified, thus, hydrogen concentration might play an important role in those metals. Although for *in situ* catalysis, this is essentially part of catalytic nature, the author needs to point it out somehow.

For the results and discussion part, the author compared two kinetics, weight normalized and surface area normalized. These two normalized reaction kinetics are based on the experimental proved fact that k [$h^{-1}$] is proportional to the weight or the surface area of the active metals. It seems reasonable for the author to investigate these two normalized kinetics together. However, several questions and doubts are drawn in this part.

1. Whether it is worthwhile a whole page to discuss the different view angles of these two kinetic.

The author spent half of the results and discussion part in two kinds of kinetics (and the other half on explaining the failure of detecting ammonium), the only difference of which is the way of normalization of $k_{obs}$. Doubt shall be casted on the motivation of this comparison. Because the author stated in the objectives of this study: determine the activity of various metal catalysts, and since weight normalized kinetics is more conventional and more practical, different normalizing kinetics should not be emphasized too much. It is better for the author to bring up this discussion as a supplemental part of metal weight normalized kinetics rather than a major part.

2. The mechanism for Fe-Ni bimetallic catalyst is ambiguous.

Zero-valence Fe based metal serves as *in situ* catalyst which consumes itself, while Ni needs dihydrogen to obtain atomic hydrogen. Needless to say, the mechanisms for Fe based and Pd or Ni based metals are totally different. However, in the case of Fe-Ni, the author seems to have assumed that only Fe contributes to the catalytic capability, thus $k_{obs}$ is only normalized by the weight or surface of Fe rather than Ni.

From table 3, the weight normalized reaction rates for Fe, Ni, Fe-Ni are 0.13±0.09, 8.3±2.9, 0.65±0.01, respectively. We can see clearly that the catalytic capability of Fe-Ni is among that of Fe and Ni. If taken into account of the role that Ni plays in Fe-Ni not as non-catalytic promoter for Fe, but as catalyst itself, then it is rather reasonable to have a reaction rate between Fe and Ni. However, Ni was not treated as active metal in Fe-Ni by the author, especially when Ni constitutes up to 42% by weight in the bimetallic catalysts. In that way, we cannot know exactly whether Ni and Fe catalyze the reaction individually, or have conjunction function. The suggestion is to use little amount of Ni in the Fe-Ni bimetallic catalyst, for instance, 0.25% by weight, as has been studied by Gui *et al.* .

3. Explanation on the function Cu serves in Pd-Cu is unpersuasive.

The author proposed a mechanism for the function of Cu in Pd-Cu bimetallic catalyst from the significant improvement (6-fold) of Pd catalyst performance with the addition of a small amount of Cu. Cu is proposed to rapidly activate NDMA which is then reduced by atomic hydrogen at the Pd surface.

The reason for making this conclusion stems from the observation that Cu cannot react with NDMA, but can promote Pd's catalytic capability. Pd is thought to activate hydrogen on the surface, which is then passed to reduce NDMA that has been activated on Cu surface with a speed faster than Pd could provide. However, since Pd itself can activate NDMA, why not it is that Cu can help Pd better activate NDMA but rather Cu can better activate NDMA itself. Since, as mentioned by the author, the activation of NDMA on Cu's surface needs further investigation, and Pd has already been proved to be able to activate NDMA, it should be reasonable to make the hypothesis that Cu can promote the activation capability of Pd for NDMA.

Apart from the problem with ammonium detection issues, the author provided a very solid prove of the mechanism in scheme 1. The concentration profiles of NDMA and DMA coincide with the pseudo $1^{st}$ order reaction assumption very well, and the carbon mass balance clearly shows that there is no intermediate during the reaction. However, for N-group, more work needed to be done.

**Conclusion:**

In this paper, several metal and bi-metal catalysts are tested on NDMA, a potential pollutant in water which is hard and less cost-efficient to remove by traditional methods. Carefully designed batch experiments are carried out for these metals. Weight normalized and surface normalized kinetics show that Pd, Pd-Cu, Ni are robust and high in reduction rate constant. No intermediate is detected during the reaction, and a hypothesized reaction mechanism is proposed based on the carbon mass balance data.

In general, this is a mediocre paper. Although lots of experiments have been carried out to test different metals, and there is abundant data, the discussion in terms of two different kinetics is weak and unpersuasive in some aspects. Too much uncompleted work weakens the discussion on mechanism. More needs to be done regarding the product detection and metal surface characterization.